# Smart Model Routing for Order-of-Magnitude LLM Cost Reduction

*A Technical Analysis of Cost-Aware Inference Routing for Large Language Models*

**Hesham Elbakoury**
Independent Researcher
helbakoury@gmail.com

**Deep Shah**
Independent Researcher
deepshah146@gmail.com

**Ankur Sharma**
Independent Researcher
San Francisco, USA
ankur.sharma@ocproject.net

## Abstract

Frontier large language models (LLMs) deliver the strongest response quality but carry per-token prices that can exceed those of smaller, task-adequate models by more than an order of magnitude. Smart model routing addresses this mismatch by dynamically assigning each incoming query to the cheapest model expected to answer it adequately, rather than sending every request to the most capable (and most expensive) endpoint. This paper develops a formal cost-quality optimization framework for routing — including a derived optimal threshold-policy structure, an explicit distinction between calibration and discrimination, a cascade breakeven condition, and a sensitivity analysis over price ratio and traffic skew — situates that framework against the principal routing architectures reported in the literature and the practical signals production systems use to estimate query difficulty before answering, and works through a numerically grounded example showing how a calibrated router can reduce blended inference cost by roughly an order of magnitude relative to an always-frontier deployment under specific, stated pricing and traffic assumptions, reporting both the gross reduction from routing policy alone and the net reduction once the router's own overhead is priced in. We situate this figure against the empirical range reported by published routing systems (roughly 2x to matching frontier-model accuracy at one-fifth its cost, depending on task skew and quality tolerance), and close with the practical risks — under- and over-routing, an aggregate quality constraint that can mask segment-level harm, calibration without discrimination, calibration drift, prompt injection into routing signals, and operational failure modes such as tier outages and multi-provider price variation — that determine whether the savings materialize in production. The analysis is primarily analytical, complemented by an empirical evaluation of the cascade framework on GSM8K reported in Section 5, validated as reported in Section 5.2; extending the evaluation across tasks and model families is future work.

## 1. Introduction

The commercial LLM market has bifurcated into two broad tiers: a small number of frontier models that lead on open-ended reasoning, coding, and instruction-following, and a much larger population of small and mid-size models that trail the frontier on the hardest tasks but are frequently indistinguishable from it on the easy and moderate ones that make up the bulk of real production traffic. Because API pricing scales with model capability, this quality gap is mirrored by a cost gap of one to two orders of magnitude between the cheapest and most expensive endpoints from a single provider's own lineup.

Most production systems still route every request to a single, statically chosen model. This is operationally simple, but it is economically wasteful whenever the chosen model is more capable — and more expensive — than a request actually requires. Smart model routing is the practice of inserting a lightweight decision layer between the application and the model backends that predicts, for each incoming query, which available model is the cheapest one likely to produce an acceptable answer, and sends the query there. When a large fraction of traffic is "easy" relative to the frontier model's capability ceiling, routing can concentrate spend on the hard tail of the distribution and route the easy majority to far cheaper models, without a proportional loss in aggregate quality.

This paper makes three contributions. First, it formalizes the routing decision as a constrained cost-minimization problem, derives the algebraic conditions under which large cost reductions are achievable, and shows that the threshold rules used throughout the taxonomy in Section 4 are instances of a single derivable optimal policy (Section 3). Second, it organizes the published routing literature into a small number of architectural families and explains the cost-quality mechanism each one exploits (Section 4). Third, it works through a concrete, numerically transparent example — using a hypothetical pricing schedule representative of September 2026 API rates and pricing the router's own overhead explicitly rather than assuming it away — that shows how a reduction of roughly an order of magnitude in blended inference cost arises from a stated query-difficulty skew (Sections 5–6), and discusses where the savings are fragile in practice, including a sensitivity analysis over price ratio and traffic skew and an explicit accounting of what this analysis does not establish (Sections 3.4, 7–8).

Throughout, we are careful to distinguish between reductions reported in specific benchmark evaluations, which vary widely by task mix and quality bar, and the illustrative 10x figure in the title, which is a representative outcome for the high-traffic-skew regime common in production support, retrieval, and classification workloads, achieved in this paper's own worked example only after its overhead is priced in — not a universal constant, and not a result this paper claims to have measured on real traffic.

## 2. Background: The Cost Structure of LLM Inference

Commercial LLM APIs bill per token, with separate (and asymmetric) rates for input and output tokens, and the rate scales with the provider's assessment of the model's marginal capability and serving cost. This paper uses a hypothetical three-tier pricing schedule, chosen to fall within the range commercial providers charted as of September 2026 without being tied to any specific named model or endpoint: a frontier tier at US$5 per million input tokens and US$30 per million output tokens, a mid tier at US$2 / US$12, and a small, latency-optimized tier at US$0.20 / US$1.20 per million tokens — roughly a 25x spread between the cheapest and most expensive tier for input tokens, and a similar spread for output. Real per-token rates move quickly and vary by provider, so these figures should be read as illustrative inputs to a cost model rather than a quote for any actual product, and every downstream number in this paper that depends on them (Sections 3.4 and 6 especially) should be re-derived with current, named pricing before being used for a real purchasing decision.

The central empirical fact that routing exploits is that this price spread is much larger than the corresponding quality spread on the majority of real queries. Benchmarks that stratify queries by difficulty consistently find that small and mid-size models match frontier-model output on straightforward classification, extraction, short-form Q&A, and templated generation tasks, and fall behind primarily on multi-step reasoning, ambiguous or adversarial prompts, and domain-specialist tasks outside their training distribution. Because production traffic for most applications is dominated by the easy and moderate end of this distribution, a policy that reserves the frontier model for the hard tail can, in principle, track frontier-level aggregate quality at a small fraction of frontier-level cost.

Two independent levers therefore determine how much a routing system can save: (i) the price ratio between the cheapest and most expensive usable model, and (ii) the fraction of traffic that genuinely requires the expensive model to meet the target quality bar. Section 3 makes this relationship explicit.

A useful intuition pump: sending every request to a frontier model is akin to hiring a senior architect to also rename files, sort tickets, and reformat dates. The architect can technically do all of this, but doing so is a poor allocation of a scarce and expensive resource. Smart routing is the inference-time equivalent of giving the routine work to a junior generalist and reserving the architect's time for the problems that actually need that level of judgment.

## 2.1 Application-Level Routing vs. Internal Model Routing

It is worth distinguishing the routing this paper discusses from a superficially similar mechanism inside some model architectures themselves. Sparsely-gated mixture-of-experts (MoE) models (Shazeer et al., 2017 [8]) route each token to a subset of expert sub-networks inside a single forward pass, as a way of increasing model capacity without a proportional increase in compute per token. That routing is internal, differentiable, trained jointly with the rest of the network, and invisible to the application calling the model. The routing this paper is concerned with is external: it happens between the application and one or more independently priced, independently served models, is typically not differentiable end-to-end with any of those models, and its explicit purpose is cost allocation across billing boundaries rather than capacity allocation within one. The two mechanisms share a name and a general shape but solve different problems at different layers of the stack.

Application-level routing is also usefully contrasted with conventional load balancing. A load balancer distributes traffic across largely interchangeable server replicas to manage throughput and availability; a model router distributes traffic across backends that differ substantially in capability, cost, and failure modes, and its objective is Equation (1)'s cost-quality trade-off rather than throughput alone. The analogy extends naturally to tiered memory: just as a memory hierarchy keeps hot data in SRAM/HBM and spills colder data to CXL-attached or pooled DRAM — paying the slower tier's latency and bandwidth cost only for the traffic that misses the fast tier — a model router keeps the bulk of traffic on the cheap tier and pays the expensive tier's price only for the queries the cheap tier cannot serve. In both cases the design problem is the same: size the fast tier's hit rate against the cost ratio of the tiers.

# 3. Formalizing the Routing Problem

## 3.1 Cost-Constrained and Quality-Constrained Formulations

Let x denote an incoming query drawn from a distribution D over the application's traffic, and let there be K available models ("tiers") indexed i = 1, ..., K, each with a per-query cost c_i(x) and an expected response quality q_i(x) ∈ [0, 1] as assessed against a task-appropriate quality metric (e.g., a preference-model score, an exact-match rate, or a rubric-graded score). A routing policy π assigns each query to a (possibly randomized) distribution over tiers, π_i(x) = P(route x → tier i). The router's design goal is most naturally stated as a quality-constrained cost minimization:

$$\pi^* = \operatorname{argmin}_{\pi} \mathbb{E}_{x \sim D}\left[\sum_{i=1}^{K} \pi_i(x)\, c_i(x)\right] \tag{1}$$

subject to the aggregate quality constraint

$$\mathbb{E}_{x \sim D}\left[\sum_{i=1}^{K} \pi_i(x)\, q_i(x)\right] \geq Q_{\min} \tag{2}$$

where Q_min is the minimum acceptable aggregate quality, typically set relative to the frontier model's own quality on the same traffic (e.g., 95–99% of frontier quality). Equations (1)–(2) make explicit that routing is not free: any router that is not perfectly calibrated trades some quality for cost savings, and the size of that trade is a design choice encoded in Q_min.

### 3.1.1 A Hidden Flaw: Aggregate Constraints Can Hide Segment-Level Harm

Equation (2) constrains only the traffic-wide average quality, and an average is a poor guardrail against concentrated harm. A policy can satisfy Equation (2) exactly while driving quality on an identifiable subset of queries — a minority language, a query pattern more common among a particular user segment, or a long-tail intent underrepresented in the router's training data — to near zero, provided the majority segment's quality is high enough to keep the average above Q_min. This is a real risk whenever the difficulty estimator's errors correlate with segment membership rather than being spread uniformly across traffic, which is the common case in practice: minority-language queries, for instance, are simultaneously more likely to be misjudged by an embedding-based estimator trained mostly on majority-language data and more likely to be quietly absorbed into an average that still looks acceptable overall.

A more defensible formulation replaces the single aggregate constraint with a constraint per traffic segment s drawn from a partition S of the traffic distribution (by language, geography, declared use case, or any other axis the deployer has reason to monitor):

$$\mathbb{E}_{x \sim D_s}\left[\sum_{i=1}^{K} \pi_i(x)\, q_i(x)\right] \geq Q_{\text{min},\, s} \quad \forall\, s \in \mathcal{S} \tag{2′}$$

Equation (2′) is strictly more conservative than Equation (2) — provided the segment floors are chosen so that their traffic-weighted average is at least Q_min, satisfying (2′) for every segment implies satisfying the aggregate version, but not the reverse — and it directly prevents the failure mode above by giving each segment its own floor Q_min,s rather than sharing one pooled budget. Its cost is operational: it requires the deployer to define the segments worth protecting and to measure quality within each one, which the purely aggregate formulation does not. Where the floors do not average to Q_min, the aggregate constraint should be retained alongside the segment constraints rather than replaced by them. Note, too, that a segment-average floor only pushes the offsetting problem one level down: within a segment, low-quality requests can still be masked by high-quality ones, so the choice of segment granularity is itself doing real work.

The stakes are easiest to see in a stark version of the example: suppose the frontier model answers every request correctly, the cheap model is correct on 99% of traffic overall, but that remaining 1% is not spread evenly — it is concentrated in a single critical subgroup (say, a specific medication-interaction question type) on which the cheap model fails 100% of the time. Sending everything to the cheap model satisfies an aggregate target of "99% of frontier quality" exactly, while failing that subgroup completely. This is not a hypothetical edge case; it is the direct algebraic consequence of optimizing Equation (2) alone. The hard safety-override rules described in Section 4.1 are best understood as a limiting case of Equation (2′): for a protected segment s, setting the floor from the frontier tier's segment-level quality — for example Q_min,s equal to the average of q_C(x) over the segment — approximates the override behavior Section 4.1 describes informally. But a quality floor alone does not force frontier routing: a cheaper tier that matches frontier quality on that segment would satisfy the floor without ever calling the frontier tier, and a per-request quantity such as q_C(x) cannot literally equal a segment-level constant. Where frontier execution is genuinely mandatory, the requirement is clearer when encoded directly as a routing restriction, π_C(x) = 1 for x in the protected segment, rather than smuggled through a quality floor. Equation (2) by itself supplies no such connection — it is compatible with an override policy, but does not imply or enforce one.

A complementary refinement addresses a different asymmetry: under-routing (sending a hard query to a tier that fails it) and over-routing (sending an easy query to a tier that succeeds at unnecessary cost) are not equally costly, yet Equation (2) treats a unit of quality shortfall the same regardless of which direction it comes from. A router training objective that reflects this asymmetry penalizes shortfalls directly rather than relying solely on the post-hoc traffic-share monitoring described in Section 7.1:

$$\ell(x) = c_{\pi(x)}(x) + \gamma \cdot \max(0,\ q_{\min}(x) - q_{\pi(x)}(x)), \quad \gamma > 0$$

where q_min(x) is the query-specific quality floor and γ prices a unit of quality shortfall in the same currency as the cost term — so γ carries units of cost per unit of quality, and its value must be chosen with those units in mind. The penalty excludes an inadequate choice only when γ exceeds the cost saved per unit of quality shortfall relative to an adequate alternative; the bare condition γ > 1 guarantees nothing by itself. For example, if A costs $0.0006 with quality 0.949 while C costs $0.015 with quality 0.950 and the floor is 0.950, then even γ = 2 gives A a penalized loss of $0.0026 — still far below C's $0.015 — so the inadequate tier wins. The objective is asymmetric by construction, since over-routing (choosing an unnecessarily expensive tier) incurs only the ordinary cost term and no penalty, but it expresses a soft preference, not a guarantee: where adequacy is mandatory it should be written as a hard constraint rather than entrusted to γ. Sections 4–6 of this paper work with the simpler aggregate constraint in Equation (2) throughout, both because it is what the cited literature reports against and because it keeps the worked example in Section 6 tractable; a deployment that adopts Equation (2′) or the penalty in Equation (2b) should expect the achievable cost-reduction factor R to be somewhat lower than Equation (4) predicts, since both refinements rule out policies the unconstrained aggregate optimum would otherwise use.

A related, but separate, imprecision concerns how Q_min is expressed. Stating a target as "95–99% of frontier quality" is only meaningful if the underlying score has a stable zero point and interval scale; shifting the scale (e.g., rescaling a preference score) can change which policies clear the threshold without changing the actual ordering of policies by quality. Even restricting attention to a metric where percentages are well-behaved, such as accuracy, a percentage-of-score target can still understate the degradation it permits: if the frontier model is 99% accurate, retaining 95% of that score permits only 94.05% accuracy — but that is a rise in the error rate from 1% to 5.95%, nearly sixfold. A quality tolerance is better expressed directly in the units that matter for the decision at hand — percentage-point accuracy loss, or excess expected loss relative to the frontier baseline — than as a bare percentage of an arbitrarily scaled score.

### 3.1.2 Calibration Is Not Discrimination

Section 7.1 relies on a "calibrated router" as the mechanism that keeps quality loss bounded, but calibration and discrimination answer different questions and neither implies the other. A router is calibrated if its predicted success probabilities match observed frequencies; it is discriminating if it can actually tell, request by request, which specific queries the cheap tier will get right. A router can be perfectly calibrated while carrying zero discriminating power, and in that regime the cost-reduction machinery in Section 3.2 collapses.

Concretely: suppose every request looks identical to the router — there is no feature that distinguishes one request from another — with the cheap model correct 90% of the time and the frontier model correct 99% of the time on this traffic. Predicting 0.90 for every request is perfectly calibrated (the predicted probability exactly matches the observed frequency), but the router has no way to identify which 90% the cheap model will get right, because by assumption there is nothing to distinguish them. If the deployment target is 98% average correctness, the only lever available is the mixing fraction φ itself: solving (1 − φ)(0.90) + φ(0.99) = 0.98 gives φ ≈ 88.9% — at least 88.9% of traffic must go to the frontier model regardless of how well-calibrated the router is. At a 25x price ratio, Equation (4) then gives R(0.889) ≈ 1.12x: essentially no savings, despite a router that is, by the calibration criterion alone, functioning exactly as intended. Conversely, rescaling a router's output scores while preserving their rank order and re-tuning the threshold to match can leave every actual routing decision — and therefore every dollar of savings — completely unchanged, which shows calibration quality and routing usefulness can also vary independently in the other direction. A reported high average success rate is consequently not evidence that an identifiable, safely routable cheap-tier subset exists; what matters is discrimination — the router's ability to separate easy requests from hard ones — and that should be reported and evaluated separately from calibration, for instance by showing realized quality as a function of the cheap-tier share the router is asked to hit, holding the underlying model pair fixed.

## 3.2 The Two-Tier Case and the Cost-Reduction Factor

The mechanics are clearest in the two-tier case — one cheap model A and one expensive model C — which also underlies most cascade and threshold-based routers in practice (Ding et al., 2024 [2]; Chen et al., 2023 [1]). Let φ ∈ [0, 1] denote the fraction of traffic the router sends to the expensive tier C, and let c_A and c_C be the (blended, per-query) costs of the two tiers. The blended cost under routing is

$$C_{\text{routed}} = \varphi\, c_C + (1 - \varphi)\, c_A$$

(3)

and the cost-reduction factor relative to an always-expensive baseline (φ = 1) is

$$R(\varphi) = \frac{c_C}{\varphi\, c_C + (1-\varphi)\, c_A} = \frac{\rho}{\varphi\rho + (1-\varphi)}$$

(4)

where $\rho \equiv c_C / c_A$ is the per-query price ratio between the two tiers. Equation (4) is the key structural result: as $\varphi \to 0$, $R(\varphi) \to \rho$, so the maximum achievable saving is bounded by the price ratio between the cheapest and most expensive usable tier, and it is realized only as the fraction of traffic requiring the expensive tier shrinks toward zero. For the September 2026 pricing used in Section 6 ($\rho \approx 25$ for a 600-input/400-output-token query mix), routing 5% of traffic to the frontier tier already yields $R(0.05) \approx 11.4$x, and routing 2% yields $R(0.02) \approx 16.9$x — before accounting for the intermediate mid-tier that a real three-tier deployment would also use, which lowers the achieved multiple somewhat, as shown in Section 6.

Equation (4) uses $\varphi$ as a raw fraction of requests, which is only equivalent to a cost-share when $c_A$ and $c_C$ are constant across the requests actually assigned to each tier. This need not hold, and the general formulation in Section 3.1 (which allows $c_i(x)$ to depend on the query) already flags the possibility — but it is worth making the failure mode concrete, because the more common real-world case works against the routing operator, not for it: harder requests are frequently also longer and more expensive ones. Suppose the 2% of requests routed to the frontier tier use, on average, 100x as many tokens as the other 98%, while the frontier tier's per-token price is still 25x the small tier's. Naively plugging $\varphi = 0.02$ and $\rho = 25$ into Equation (4) gives $R(0.02) \approx 16.9$x, but the actual reduction — weighting each tier's contribution by the tokens it actually processes — is only

$R = (0.98 + 0.02 \times 100) / (0.98 / 25 + 0.02 \times 100) \approx 1.46\times$,

more than an order of magnitude below the naive estimate. The general fix is to weight by relative per-request cost rather than raw request count. Writing $\tau$ for the ratio of average token count (or, more generally, total per-request cost at matched pricing) between frontier-routed and small-routed traffic, the corrected formula is

$$R(\varphi, \tau) = \frac{(1-\varphi) + \varphi\tau}{(1-\varphi)/\rho + \varphi\tau}$$

(5)

which reduces to Equation (4) exactly when $\tau = 1$ (frontier-routed and small-routed requests cost the same on average) and degrades gracefully otherwise. Because $\tau \neq 1$ is the practically relevant case whenever query difficulty correlates with length — a very plausible correlation for open-ended reasoning or multi-document synthesis tasks — any reported R should either confirm $\tau \approx 1$ on the traffic in question or use Equation (5) with a measured $\tau$ rather than the simpler Equation (4).

### 3.3 Cascade Routing

A cascade router (Chen, Zaharia, and Zou, 2023 [1]) does not decide up front which tier to use; instead it calls the cheapest tier first, evaluates the response against a confidence or agreement criterion, and escalates to the next tier only on failure. For a K-tier cascade with per-tier acceptance probability a_i (the probability that tier i's answer is accepted given that the query reached tier i), the query reaches tier i only if it was rejected at every earlier tier, so the expected cost is

$$\mathbb{E}[C_{\text{cascade}}] = \sum_{i=1}^{K} c_i \, P_i \,, \quad P_1 = 1 \,, \quad P_i = P_{i-1}(1 - a_{i-1}) \tag{6}$$

where P_i, defined recursively above, is the probability a query reaches tier i — that is, the query reaches tier i only if it was rejected at every earlier tier. Cascades save cost whenever the cheap tier's acceptance rate a_1 is high, because the expensive tier is then invoked on only the (1 − a_1) fraction of traffic that the cheap tier itself flags as uncertain — the escalation decision is made by the system itself rather than by a separately trained classifier, which is the main practical advantage over the classifier-routing approach of Section 4.3, at the cost of always paying for at least the cheap-tier call on every query.

Equation (6) is a cost model only, and it is tempting — but wrong — to reuse the per-tier quality values q_i(x) from Section 3.1 unchanged to describe a cascade's quality, because a cascade's final answer is not a random draw from tier i's unconditional output; it is drawn from tier i's output conditional on having been accepted by the checker. These can differ enormously, and in either direction. Consider two models that are each independently correct with probability 0.5. Any one-shot mixture of the two — however the mixing weights are chosen — stays at exactly 0.5 correct, since a weighted average of two 0.5's is 0.5. But a cascade with a checker that perfectly identifies whether the first model's specific answer is correct, escalating only when it is not, reaches 0.5 + 0.5 × 0.5 = 0.75: strictly better than either model alone, and better than any one-shot mixture of them. This is not a claim that a perfect checker exists or is free to build — it is a demonstration that a cascade's checker is not just a cost-control device, it is also an information source that changes what quality is achievable at all. Writing a_i for the acceptance rate at tier i and q_i^acc for the quality of tier i's answer conditional on that answer being accepted (as opposed to the unconditional q_i used in Section 3.1), the correct two-tier cascade quality is

Q_cascade = a_1 · q_A^acc + (1 − a_1) · q_B^rej, where q_B^rej is tier B's quality conditional on the request having been rejected at A — the symmetric counterpart of q_A^acc, since B never sees a random draw of traffic, only the rejected residual.

which collapses to the toy example above when a_1 = 0.5, q_A^acc = 1 (the checker accepts only correct answers), and q_B^rej = 0.5. The gap between these conditional qualities — q_i^acc on the accepted side, q_i^rej on the rejected side — and the unconditional q_i is exactly what Equation (6) leaves unmodeled, and it is governed by the checker's informativeness, not by the acceptance rate a_i alone — a checker with a high a_1 but poor informativeness (accepting almost everything, correct or not) drives cost down per Equation (6) while q_A^acc stays close to the unconditional q_A, capturing little of the 0.75-style gain the toy example illustrates. Concretely, if half the traffic is easy and half hard, A answers every easy query correctly, a perfect checker accepts exactly those, and B answers every easy query correctly but only 20% of hard ones, the naive unconditional form gives 0.5 + 0.5 × 0.6 = 0.80 while the true cascade quality is 0.5 + 0.5 × 0.2 = 0.60 — because B only ever sees the hard requests. For longer cascades, each tier's contribution is conditioned on the full rejection history that delivered the request to it.

The same conditioning logic extends to any number of tiers. Writing P_i for the probability a query reaches tier i — with P_1 = 1 and P_i = P_{i−1}·(1 − a_{i−1}) — and taking the terminal tier to always accept (a_K = 1), the exact expected cascade quality is:

$$\mathbb{E}[Q_{\text{cascade}}] = \sum_{i=1}^{K} P_i \cdot a_i \cdot q_i^{\text{acc}}$$

The terminal tier's q_K^acc is then simply its quality conditional on the full rejection history — what the two-tier formula calls q_B^rej — so this K-tier form reduces exactly to the two-tier equation when K = 2.
That last clause is not a minor caveat: a cascade is not automatically cheaper than simply calling the escalation tier directly. For the two-tier case (cheap tier A escalating to tier B), Equation (6) reduces to E[C] = c_A + (1 − a_1)c_B, and comparing this to the cost of always calling B directly gives the breakeven condition

$$a_1 > \frac{c_A}{c_B}$$

If tier A's acceptance rate falls below the cost ratio c_A/c_B — for instance, because the confidence check is too conservative, or because traffic drifts toward queries A handles poorly — adding stage A costs more than skipping straight to B, and it does so while also adding the latency of two sequential calls on every escalated query. This is a local statement, however, about whether one specific stage is worth adding given the cost of the stage immediately after it — it is not a claim about whether a full multi-stage cascade beats its own strongest-model baseline system-wide, and the two questions can have opposite answers. Take a three-tier cascade with per-tier costs 1, 2, and 100, where stage A never accepts (a_1 = 0) and stage B accepts 99% of what reaches it. Stage A violates its own local breakeven condition by construction (a_1 = 0 fails a_1 > c_A/c_B for any positive ratio) and is, locally, pure waste — yet the full cascade costs 1 + 2 + 0.01×100 = 4, still 25x cheaper than the cost-100 baseline of always calling the top tier directly. A locally wasteful stage does not imply a system-level loss, and a locally justified stage does not by itself guarantee one either, since the local condition never accounts for what happens beyond the very next tier. System-level savings claims need a comparison of complete path costs across the whole cascade, or an explicit with-versus-without-this-stage comparison, not an appeal to the local condition alone. Cascade designs also need the same ongoing acceptance-rate monitoring that Section 7.1 recommends for classifier-based routers; a cascade that looked cost-effective at launch can quietly become a net loss as a_1 drifts down. For multi-stage cascades (K > 2) the local pairwise conditions are neither sufficient for global cost optimality — every stage can satisfy its own a_i > c_i/c_{i+1} check while a different subset of stages is cheaper overall — nor a valid pruning rule. Failing the adjacent-tier check does not by itself justify deleting a stage, because the check prices the saving from an accepted request at exactly one downstream call, whereas in a longer cascade an acceptance can avert several downstream calls, including an expensive escalation triggered when a later stage's checker rejects traffic the deleted stage would have absorbed. The condition $a_A > c_A/c_B$ remains valid for the two-stage comparison in which B terminates the request; for $K > 2$ the correct test compares the complete expected path costs with and without the stage, with quality evaluated separately. A counterexample on the same 1/2/100 costs shows the failure concretely. Let traffic divide into groups of 10%, 80%, and 10%, where tier A answers and accepts only the first group, tier B answers the first two groups correctly but its checker accepts only the middle group, and tier C always answers correctly, with checker overhead ignored. With stage A in place the expected cost is $1 + 0.90 \times 2 + 0.10 \times 100 = 12.8$; without it, $2 + 0.20 \times 100 = 22$. Stage A fails its local check ($a_A = 0.10 < c_A/c_B = 1/2$) and B matches A's answer quality on the traffic A accepted, and final-answer quality is 100% in both systems, yet deleting A raises expected cost by 71.9%: A's accepted traffic, once re-routed, is rejected by B's checker and falls through to the $100 tier. No quality proviso repairs this, because the error is in the cost accounting rather than the quality comparison. The 1/2/100 example above marks the boundary between the two regimes: stage A fails its local check and the (B→C) sub-cascade is indeed cheaper (3 rather than 4 at identical quality), but only because A accepted nothing, so deleting it re-routed no traffic — the degenerate case in which the local check and the full path-cost comparison coincide.

Cascades work best when the accept/reject decision can be made by an automatic, objective check rather than a subjective quality judgment — but an objective check is not automatically a reliable one, and the gap between the two is exactly the $q_i^{acc}$ term above. Structured-extraction tasks illustrate the risk as clearly as the opportunity: if a small model is asked to pull a date, an account number, and a total from an invoice, verifying that all three fields are present, the date parses, and the amount is numeric confirms only that the answer is well-formed, not that it is correct — a wrong account number, an invented date, or an incorrect total can pass every one of those checks while being substantively wrong. The quantity that actually determines $q_A^{acc}$ is the checker's false-accept rate: the probability an incorrect answer is nonetheless accepted. A checker that accepts almost everything produces excellent cost numbers under Equation (6) while returning many wrong answers with a $q_A^{acc}$ barely above the unconditional $q_A$. Deployers should define precisely what each check does and does not verify, measure the error rate specifically among accepted answers (not just overall), and where feasible verify extracted values against their source rather than only their format. Code generation offers a similar hook — generated code can be run against a test suite — but the same distinction applies: passing the supplied tests is not the same claim as being correct beyond those tests, and a false-accept rate exists here too, just measured against a different notion of ground truth. Where no automatic check exists — judging whether a business recommendation is sound, or whether an explanation is genuinely clear — cascades must fall back to a separate evaluator model or a classifier-based approach, reintroducing the calibration problem cascades are otherwise able to partially sidestep, and even where an objective check does exist, it validates only what it was designed to detect, not the acceptance policy as a whole.

Finally, the cascade's accept/reject decision is a different optimization problem from the ex-ante routing choice of Section 3.5, even though the two share the same marginal-value shape. Once tier A has answered, its cost is sunk: the decision to continue to B compares the expected quality gain — conditioned on everything observed so far, namely A's answer and the checker's signal, summarized as history h — against B's full additional cost, including any remaining checker costs, rather than against the cost difference c_C − c_A that appears in Equation (8):

$$\lambda\left[\mathbb{E}(q_B \mid h) - \mathbb{E}(q_A \mid h)\right] \geq c_B + c_{\text{check}}$$

In words: escalate when the shadow-priced expected quality gain, updated on the observed history, covers the incremental cost of the next stage. Equation (8) remains the right description of the direct-routing decision — which tier to try first — while this rule governs the interim stop-or-continue decision inside a cascade.

## 3.4 Worked Numerical Bound

Combining Equation (4) with published price ratios gives a quick sanity check on any claimed routing saving: a claim of Rx savings is only physically possible if ρ ≥ R, and it requires the fraction of traffic sent to the expensive tier to satisfy

$$\varphi \leq \frac{\rho - R}{R(\rho - 1)}$$

obtained by solving Equation (4) for φ. This bound is a useful reviewer's check, but only if ρ is measured correctly: it must be the ratio of full per-request costs, not a ratio of per-token list prices. The two diverge whenever the tiers differ in more than list price alone — a longer average response, a larger sampling count, or a more elaborate multi-step configuration on one tier all inflate its true per-request cost beyond what its per-token rate alone suggests. A published 8x gap in per-token list prices does not, by itself, rule out a genuine 50x per-request cost gap, since a frontier tier that also produces responses 8x longer than the cheap tier's — plausible if it is used for the more open-ended, harder queries per Equation (5)'s τ — compounds the two multipliers into a 64x gap in what each request actually costs. Equation (4)'s bound is correct once ρ is defined this way; it is only the shortcut of reading ρ off a pricing page, ignoring everything else that varies between tiers, that produces an unreliable check. Figure 1 visualizes Equation (4) across a grid of price ratios and frontier traffic shares, making the region in which a 10x-class claim is achievable explicit rather than asserted: the paper's own worked example in Section 6 (ρ ≈ 25, φ = 2%) sits comfortably inside that region, but the same 25x price ratio at φ = 10% falls short of 10x, and a 10x claim is unreachable at any traffic split once the price ratio falls below 10 itself. Between 10 and roughly 11 the claim survives only at very small frontier shares: at a price ratio of 10.5, Equation (4) gives 10.40x at a 0.1% frontier share, and the bound above permits a frontier share of at most about 0.53%. The roughly-11x figure is the price ratio required at a 1% frontier share, not a universal floor.

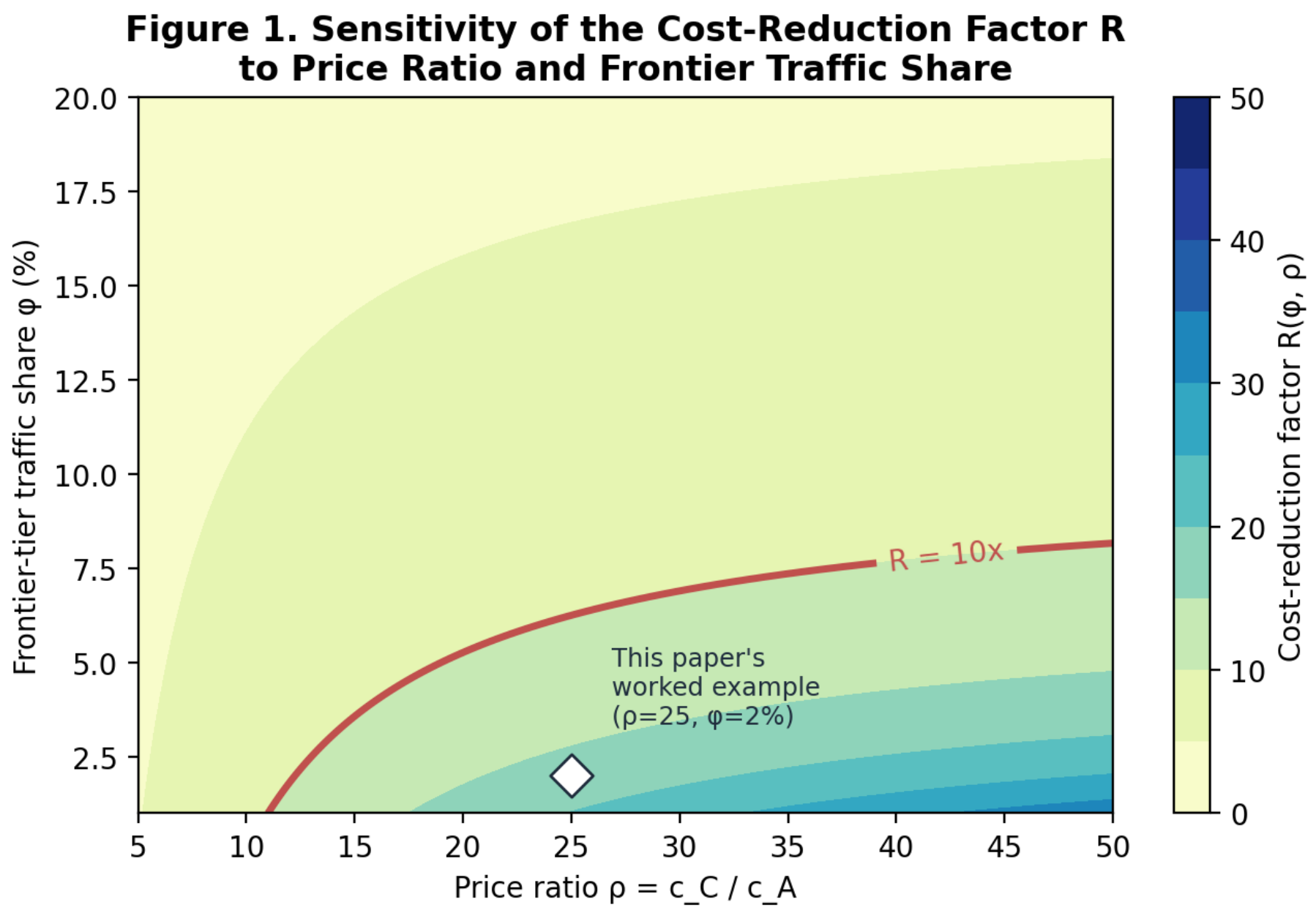

*Figure 1. Sensitivity of the cost-reduction factor R(φ, ρ) to the price ratio ρ and the frontier-tier traffic share φ, per Equation (4). The red contour marks R = 10x; the combinations that reach a 10x reduction lie below it (frontier share small enough for the price ratio), combinations above it cannot reach a 10x reduction regardless of router quality, and no combination with a price ratio below 10 reaches 10x at any traffic share.*

## 3.5 Optimal Policy Structure: A Threshold Rule

The taxonomy in Section 4 presents rule-based, predictive, cascade, and preference-trained routers as separate design choices, but the threshold-based members of that taxonomy — Hybrid LLM's tunable difficulty threshold, for instance — are approximating a common, derivable optimum. A cascade's accept/reject decision has the same marginal-value shape but is a separate, interim problem with sunk costs, derived at the end of Section 3.3. Relaxing the routing indicator π_i(x) to a soft assignment and forming the Lagrangian of Equations (1)–(2) with multiplier λ ≥ 0 for the quality constraint gives

L(π, λ) = E_x[Σ_i π_i(x)(c_i(x) − λ q_i(x))] + λ Q_min.

Because the objective is separable across queries once λ is fixed, the inner minimization can be solved independently for each x: the optimal policy simply assigns each query to whichever tier minimizes cost net of quality, valued at the shadow price λ of the constraint:

$$\pi^*(x) = \operatorname{argmin}_i \left[c_i(x) - \lambda\, q_i(x)\right] \tag{7}$$

with λ ≥ 0 chosen to satisfy the optimality conditions for the relaxed problem: complementary slackness λ·(Q(π) − Q_min) = 0, so that if the cheapest tier already clears the quality target the constraint is slack and λ = 0, and otherwise λ > 0 is tuned (for instance, by bisection) until Equation (2) binds. Where several tiers tie in the argmin at the chosen λ, the policy randomizes among the tied actions in the proportions needed to meet the quality target exactly — a deterministic threshold alone cannot always do so. For the two-tier case this reduces to a clean threshold rule on the marginal quality gained per dollar spent upgrading from A to C:

$$\text{route to } C \Longleftrightarrow \frac{q_C(x) - q_A(x)}{c_C - c_A} \geq \frac{1}{\lambda} \tag{8}$$

This is the formal justification for the direct-routing threshold rules in Section 4: Hybrid LLM's tunable difficulty threshold is, in effect, estimating the left-hand side of Equation (8) — the marginal quality return on routing a specific query to the more expensive tier — and comparing it to a cutoff set by the deployer's chosen Q_min. A cascade's accept/reject criterion follows the same marginal-value logic but applied interim, with posterior expectations and sunk costs, per the stop-or-continue rule at the end of Section 3.3. A router's "difficulty score" is therefore best understood as a learned proxy for this marginal quality-per-dollar ratio rather than an independent quantity, which is also why a difficulty estimator trained on one price ratio ρ does not automatically transfer to another: the optimal cutoff 1/λ shifts whenever the price gap $c_C - c_A$ does, even if the underlying quality functions $q_A$ and $q_C$ do not change at all. The asymmetry deserves emphasis, because it is a live implementation error, not a theoretical nicety: an illustrative check reported with the validation tests in Section 5.2 — applying Equation (8)'s subtraction at an interim cascade stage, where the first stage's cost is already sunk — systematically over-escalates and in some configurations selects escalations that lose money. Equation (8) prices a choice between two unpaid alternatives; the stop-or-continue rule prices the next stage against what has already been spent, and the two must not be interchanged. The same reading applies to RouteLLM [3]: its threshold α on a learned preference probability is a cutoff of the kind this section derives, and the strong-model call shares it reports are realized values of φ under its chosen performance target.

It is worth being precise about what Equation (8) is not, because the informal shorthand used elsewhere in this paper — "route to the cheapest model that clears a quality bar" — describes a different and generally suboptimal rule. A per-request absolute threshold accepts a query at the cheap tier whenever its predicted quality there exceeds a fixed cutoff, regardless of how much quality an upgrade would add or how much that upgrade costs; Equation (8) instead ranks upgrades by the quality gained per dollar spent and funds the best-ranked ones first, without reference to any fixed absolute cutoff. The two coincide only under special conditions on the cost and quality functions; in general they disagree, sometimes sharply. Consider two equally common request types with equal upgrade cost: type 1 goes from quality 0.79 to 0.80 on upgrade, type 2 from 0.81 to 1.00. A per-request threshold at 0.80 upgrades type 1 (0.79 is below the bar) and leaves type 2 alone (0.81 already clears it) — capturing a quality gain of just 0.01. Given a fixed budget for only one of the two upgrades, Equation (8) instead upgrades type 2, capturing a gain of 0.19, precisely because it ranks by marginal gain rather than by absolute post-upgrade level. A fixed difficulty threshold is therefore a heuristic surrogate for Equation (8), not an automatic implementation of it, and the two should not be conflated when describing what a given routing system actually optimizes; Section 4's descriptions of threshold-tuning systems refer to the surrogate unless a paper explicitly reports ranking by marginal quality-per-dollar.

# 4. Taxonomy of Smart Routing Architectures

Published routing systems differ mainly in how they estimate, for a given query, which tier is the cheapest one likely to clear the quality bar. Figure 2 shows the generic pipeline shared by most designs: a feature or difficulty estimator, a router policy that maps features to a tier (or a distribution over tiers), and — in cascade variants — a confidence check that can escalate a query the router initially under-routed.

**Figure 2. Generic Smart-Routing Inference Pipeline**

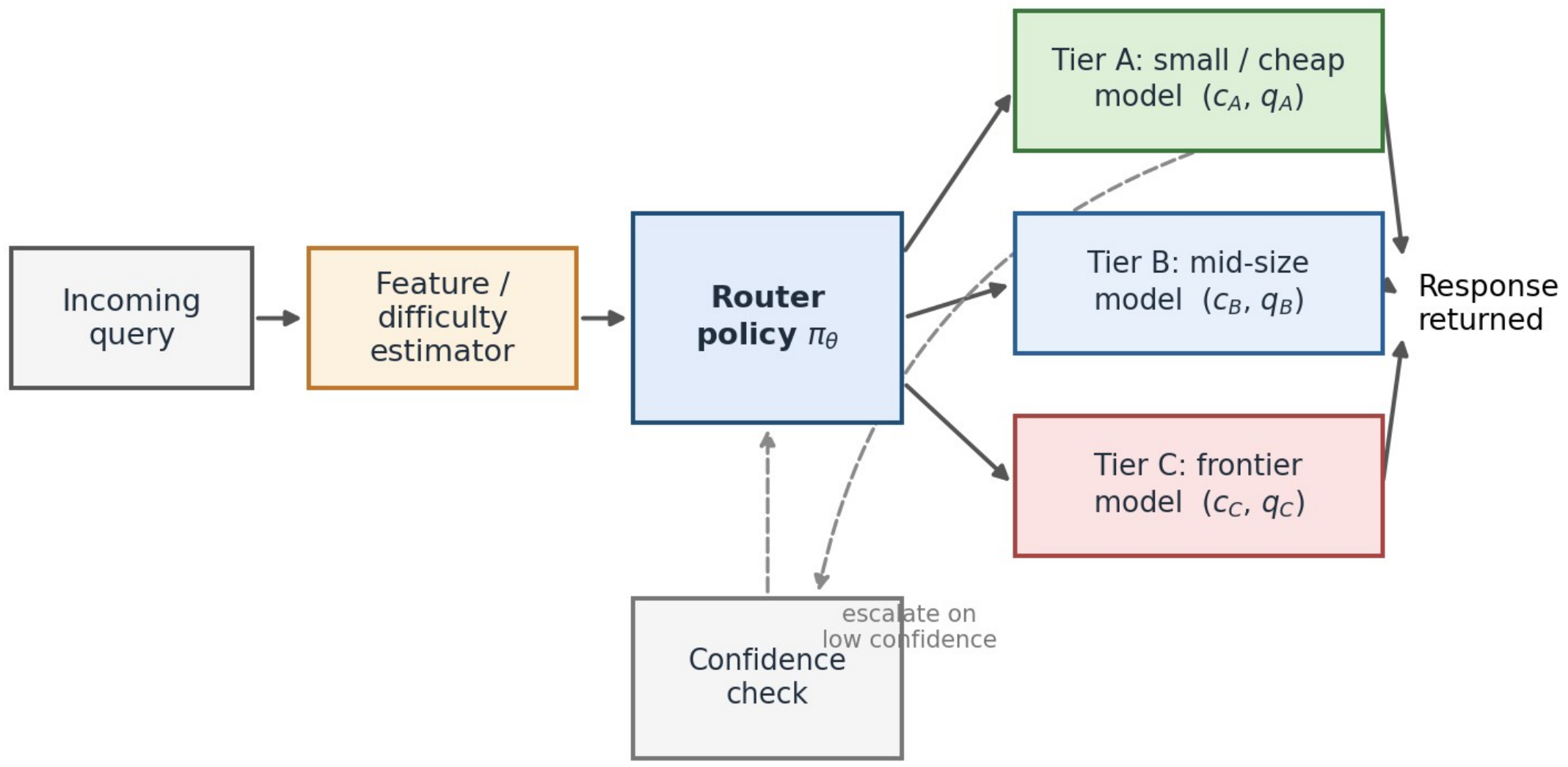

*Figure 2. Generic smart-routing inference pipeline. Solid arrows show the primary request path, ending at whichever tier the router policy selects. Dashed arrows show the optional escalation path used by cascade and hybrid designs: the confidence-check stage runs an automatic acceptance test (Section 3.3) against the response from Tier A or B, and re-invokes the router only when that test fails, sending the query to the next more capable tier rather than returning the rejected response.*

## 4.1 Signals Used to Estimate Query Difficulty

Before comparing routing architectures, it is worth being explicit about what a difficulty estimator actually conditions on, since this is where naive implementations most often go wrong. Message length is a tempting but unreliable proxy: "Is this contract valid?" is four words but may require legal expertise and extensive context, while a request to extract every email address from a long pasted document is conceptually straightforward despite its length. Production estimators instead combine several independent signals:

| Signal | What it captures | Example |
|---|---|---|
| Task type | Certain task categories carry inherently lower or higher reasoning load, largely independent of length. | Classification, extraction, translation, and formatting are typically low-reasoning; planning, debugging, mathematical proof, and reconciling conflicting documents are typically high-reasoning. |
| Risk / stakes | The cost of an inadequate answer, not just the difficulty of producing one, should influence tier selection. | Medical, legal, financial, and security-related queries are often routed to a stronger tier even when they appear linguistically simple, via a hard override rule rather than the learned estimator. |
| Context volume | How much material the model must integrate to answer correctly. | Multi-document synthesis or long conversation histories require a tier with adequate context handling, independent of the question's surface difficulty. |
| Output constraints | How structured or tightly constrained the required output is. | Populating a fixed-schema JSON object is easy; producing a technical design that must satisfy many simultaneous constraints is hard, even if the prompt is short. |

*Table 1. Representative signals combined by production difficulty estimators. No single signal is sufficient on its own; rule-based, predictive, and cascade routers in Sections 4.2–4.4 differ mainly in how they combine or bypass these signals.*

The risk signal deserves particular emphasis because it is qualitatively different from the others: it is not really an estimate of difficulty at all, but a policy decision about acceptable failure cost. Most production systems implement it as a hard override — the limiting case of Equation (2′) described in Section 3.1.1, a rule that certain query classes are always sent to the strongest available tier regardless of what a difficulty classifier concludes — precisely because a classifier's job is to estimate reasoning load, not to weigh the asymmetric cost of a wrong answer in a regulated or safety-relevant domain.

## 4.2 Rule-Based and Heuristic Tiering

The simplest routers apply fixed, human-authored rules — prompt length, presence of code fences, detected task type via keyword matching, or a user-selected "mode" — to assign a tier. These are cheap to build and fully interpretable, but because the rules are not learned from labeled outcome data, they systematically misroute queries whose difficulty does not correlate with the chosen heuristic, which caps their achievable savings well below what a calibrated router reaches for the same quality bar (illustrated numerically in Section 6).

### 4.3 Predictive (Classifier-Based) Routing

Predictive routers train a supervised model — typically a small classifier or regression model over prompt embeddings — to estimate query difficulty or the probability that a given tier will produce an acceptable answer, then apply a threshold rule. Ding et al.'s Hybrid LLM (2024) [2] is the canonical example: a DeBERTa-based router predicts, for a query, the probability that a small model's response quality (BART score) will match a large model's, and a threshold on that prediction is tuned at test time to trade cost for quality. The reported numbers are pair- and operating-point-specific rather than a single figure: at a 40% cost advantage (the fraction of queries routed to the small model), the paper's best router variant reports a 0.0% BART-score drop for the Llama-2-7B/Llama-2-13B pair, a 2.9% drop for the Llama-2-13B/GPT-3.5-turbo pair, and a 10.3% drop for the FLAN-T5-800M/Llama-2-13B pair — the drop growing with the capability gap between the two models being routed between, as Equation (8) predicts it should (a bigger quality gap between tiers means the marginal-value ranking runs out of high-return queries to upgrade sooner). Because the threshold is exposed as a single tunable knob, this family maps directly onto the $\varphi$ parameter in Equation (4), though — per the caveat in Section 3.5 — a fixed difficulty threshold implements a heuristic surrogate for the marginal-value rule in Equation (8), not that rule exactly.

A common and inexpensive variant uses a small general-purpose LLM as the classifier, prompted to output a structured difficulty judgment rather than an answer. For example:

```
Classify this request as EASY, MEDIUM, or HARD.

EASY: extraction, formatting, simple classification, or
direct rewriting.
MEDIUM: summarization, ordinary coding help, or moderate
analysis.
HARD: complex reasoning, conflicting evidence, high-risk
advice, multi-document analysis, or strict multi-step
constraints.

Return JSON: {"difficulty": ..., "risk": ...,
"recommended_model": ..., "reason": ...}
```

Because the classification prompt and its structured response are both short relative to the task itself, the amortized classification cost C_router in Equation (1) is small in isolation next to the price gap between the tiers it is deciding among. Section 6's worked example quantifies this precisely rather than assuming it away, and the classification cost alone is indeed small there — but the escalation calls triggered when a cascade's acceptance check fails are not, and together they still erode the gross reduction factor by a meaningful margin in that example. The lesson is that "small" needs a number attached to it, and the number that matters most is usually the escalation rate, not the classifier's own per-call cost. The main failure mode of this approach is that the classifier can itself misjudge a request; production deployments therefore typically layer the hard safety-override rules from Section 4.1 on top of the model's judgment, so that a classifier mistake on a flagged category cannot silently downgrade a high-stakes query.

### 4.4 Cascade Routing

As formalized in Section 3.3, cascades call the cheap tier first and escalate on failure, rather than deciding a priori. FrugalGPT (Chen, Zaharia, and Zou, 2023 [1]) is the original instantiation of this idea for commercial LLM APIs: it trains a DistilBERT-based regression model as the accept/reject scoring function g(q, a) — a learned quality estimator, not a fixed rule — and pairs it with a threshold-based "stop judger" and an optimizer that selects both the order of models in the cascade and each stage's threshold. The paper's own headline claim is up to 98% cost reduction while matching the best individual LLM's accuracy across its full evaluation suite; a more granular, concretely worked case study on its HEADLINES benchmark reports FrugalGPT matching GPT-4's accuracy at one-fifth of GPT-4's cost (and, in a separate case study, improving accuracy by roughly 1.5 points at 80% lower cost) — these are different operating points within the same paper, not competing figures, and both are the paper's own reported numbers rather than this paper's re-derivation of them. Either way, the savings trace to the scorer's ability to correctly identify which cheap-tier answers are already good enough, not merely to a lenient acceptance criterion — an important distinction given Section 3.3's point that a high acceptance rate with low checker informativeness produces the same cost numbers without the same quality.

BEST-Route (Ding et al., 2025) [7] is frequently grouped with cascades, including in an earlier draft of this paper, but its actual mechanism is closer to joint model-and-sampling-budget selection: for each query it chooses both which model to call and how many responses to sample from it (via best-of-n selection), rather than trying one tier and sequentially escalating on rejection. It does not fit the accept/reject/escalate structure Equation (6) models, and the breakeven condition in Section 3.3 does not apply to it directly — a more relevant cost driver for BEST-Route is the number of samples drawn per query, which this paper does not formalize. As reported by its authors, it reduces cost by up to 60% with an 0.80% drop in ArmoRM reward-model score at a sampling cap of 5, with a 1.59% drop reported on an out-of-distribution MT-Bench evaluation at the same 60% target — figures that, consistent with Section 7.2's caution about judge costs, do not appear to include the cost of the reward model used to select among sampled responses.

### 4.5 Preference- and Reward-Trained Routers

RouteLLM (Ong et al., 2024/2025) [3] trains routers directly on human preference data (augmented with additional labels) to predict which of a strong/weak model pair a human would prefer for a given query, rather than predicting a difficulty score as an intermediate quantity. Its own reported numbers are also operating-point- and benchmark-specific: at the paper's call-performance-threshold-50% (CPT(50%)) operating point — the minimum percentage of calls to the strong model needed to recover half the performance gap between the strong and weak models — its best router retains approximately 95% of GPT-4's score on MT Bench, 87% on GSM8K, and a comparable figure on MMLU, corresponding to reported cost-saving multiples of roughly 3.66x, 1.49x, and 1.41x respectively on the paper's own cost-normalized benchmark; the paper's headline claim of "over 2x" savings without compromising quality describes the best of these operating points, not a uniform result across benchmarks. It is also worth noting that CPT is a strong-model call-share metric, not directly a dollar-cost ratio against an always-frontier baseline — the two coincide only when converted through the same per-call price ratio used elsewhere in this paper, which RouteLLM's own appendix supplies for its specific model pair (GPT-4 and Mixtral-8x7B) but which will differ for any other pair. The authors additionally report that the trained routers transfer to new strong/weak model pairs at test time without retraining — a practically important property, since it decouples router training cost from the frequency of model upgrades — though this transfer degrades as the correlation between the training and target pairs' quality gaps weakens, per the paper's own generalization experiments.

More generally, learned routers of this kind are built from outcome data collected by sending the same representative prompts to every candidate tier and recording which tiers succeed, and how that success is measured matters as much as the collection procedure itself. Labeling from a single sampled response per tier, rather than an estimated success probability, can produce systematically wrong labels whenever tier success is itself stochastic. Suppose the small tier succeeds on a given query with independent probability 0.60 and the mid tier with probability 0.90, and the deployer's target is 0.85 expected success — only the mid tier actually clears that bar. Labeling by "which tier succeeded in this one sampled trial," the small tier succeeds and the mid tier fails 0.6 × 0.1 = 6% of the time, both succeed 0.6 × 0.9 = 54% of the time, only the mid tier succeeds 0.4 × 0.9 = 36% of the time, and neither succeeds 0.4 × 0.1 = 4% of the time. If ties are broken toward the cheaper tier (a reasonable default), "small" is the assigned label whenever the small tier succeeds regardless of the mid tier's outcome — 60% of the time — making it the majority label even though only the mid tier meets the actual target. Ding et al. [2] address exactly this problem for the two-tier case by estimating success probability from repeated sampling rather than a single draw (see Section 4.3); routers trained on single-sample labels without this correction inherit a bias toward whichever tier is stochastically "lucky" more often, not whichever tier is actually adequate, and this bias is invisible unless success is explicitly modeled as a probability rather than a one-shot outcome.

It is worth locating RouteLLM explicitly within the framework of Section 3, because it is the published system whose design most closely matches the policy structure derived there. Its decision rule — route to the strong model when a learned estimate of the strong model's win probability for the query exceeds a threshold α, and to the weak model otherwise — is a threshold rule of the form Section 3.5 analyzes: α plays the role of the cutoff induced by the quality constraint's multiplier λ, and the fraction of traffic the threshold sends to the strong model is a realized φ in the sense of Equation (4). The qualifications of Section 3.5 apply unchanged. A fixed threshold on an estimated score is the heuristic surrogate for Equation (8)'s marginal-value ranking, not the ranking itself — the same qualification Section 4.3 records for Hybrid LLM — and the cutoff must be re-set whenever the price gap between the tiers changes, even when the underlying quality functions do not. Read through Section 3.5, the authors' transfer result is the expected behavior when the marginal-value ordering of queries is preserved across model pairs, and the degradation they report as the pairs' quality-gap correlation weakens is the predicted failure mode; that a router trained once transfers at all is evidence that the threshold structure is a property of the routing problem rather than of a particular model pair. RouteLLM is also deliberately one-shot — every query is served by exactly one model — which forgoes the cascade option rather than pricing it. Section 3.3 supplies the price: whether the forgone option was worth anything on a given workload depends on the cheap tier's acceptance rate and the checker's informativeness, quantities a one-shot design never measures. Finally, its quality targets are aggregate — performance retained relative to the strong model, averaged over traffic — and therefore inherit the concern of Section 3.1.1: no per-segment floor of the Equation (2′) form is imposed, and a traffic-wide target can be met while a minority segment is systematically under-served. The router's own inference cost, small for its classifier architectures, is reported separately in the paper's Section 5.5 (Table 7) — at most about 0.4% of GPT-4's generation cost for its most expensive router — rather than netted into the headline savings figures, in contrast to the treatment in Section 6. None of this diminishes the result. RouteLLM is the strongest published evidence that a learned threshold router can retain quality at a sharply reduced strong-model call rate, and its design is, in this paper's terms, a specific and defensible point in the policy space Section 3 characterizes — one that trades the cascade option and per-segment guarantees for single-call latency and aggregate quality control.

## 4.6 Semantic and Embedding-Based Routing

A final family avoids training a dedicated difficulty classifier at all, instead embedding the incoming query and comparing it, via nearest-neighbor search, to a bank of labeled examples. This is attractive when labeled preference data is scarce, at the cost of needing an up-to-date, representative example bank and added retrieval latency. It is important to distinguish two things this technique is used for, because they are often conflated: semantic routing by intent groups queries by task type or domain — for example, recognizing that "why did you charge me twice" and "I see a duplicate payment" both belong to a billing category, and should reach a billing-specialized prompt or model regardless of the exact wording used — whereas semantic routing by difficulty attempts to estimate reasoning load from embedding similarity to previously solved queries of known difficulty. Knowing a query's topic does not, by itself, reveal whether it is a routine lookup or a complex dispute within that topic; most production systems therefore use embedding similarity for intent classification and pair it with a separate difficulty signal from Section 4.1 rather than relying on it for both.

## 4.7 Benchmarking and Evaluation Infrastructure

RouterBench (Hu et al., 2024) [4] and the survey by Varangot-Reille et al. (2025) [6] address a cross-cutting problem: routing papers historically reported savings against inconsistent baselines, model sets, and quality metrics, making cross-paper comparison unreliable. RouterBench standardizes evaluation over roughly 405,000 inference outcomes across representative LLMs and proposes cost-normalized performance metrics; the survey formalizes routing generally as the performance-cost optimization problem underlying Equations (1)–(2) above and catalogs the routing strategies published to date. Shnitzer et al. (2023) [5] independently proposed benchmark-driven router training as a way to ground router decisions in measured, rather than assumed, per-model performance.

## 4.8 Compounding Routing with Semantic Caching

Routing decides which model answers a request; it does not ask whether the request needs to be answered by a model at all. A complementary lever is semantic caching, which stores previous query-response pairs as embeddings and serves a cached response directly — bypassing every model tier — when a new query is sufficiently similar to one already answered (Bang, 2023 [9]). For applications with a meaningful rate of repeated or near-duplicate queries, such as customer-support chat, caching intercepts a share of traffic before routing logic ever runs. If h is the cache hit rate, the effective blended cost becomes $(1 - h) \cdot C_routed$, so even a modest hit rate compounds multiplicatively with whatever reduction routing already achieves — a 20% hit rate multiplies the routing reduction factor by a further $1/(1 - 0.20) = 1.25$x, and a 40% hit rate (plausible for high-repetition support workloads) by 1.67x.

This compounding claim rests on an assumption worth stating explicitly rather than leaving implicit: that a cache sitting architecturally in front of the router does not change the statistical character of the traffic the router subsequently sees. Architectural independence does not imply statistical independence — cache hits are, by construction, the more common, more repeated queries, so the traffic that reaches the router after a cache is systematically the residual, less-typical distribution, not a scaled-down copy of the original. Whatever discriminating power a router was tuned or evaluated on the full, uncached traffic distribution may not transfer to this residual distribution, and Equation (4)'s $\rho$ and $\varphi$ — along with the realized R — should therefore be evaluated on the cache-miss distribution specifically, not assumed to carry over from the pre-cache traffic. Caching also introduces its own failure mode symmetric to the router's own: a false hit — two queries similar enough in embedding space to match, but requiring materially different answers — returns a stale or wrong response with no escalation path at all, since a cache hit bypasses every tier, including whatever confidence check or classifier would otherwise have caught a bad answer. A cache lookup is also not literally free, though it is far cheaper than a model call; where this paper describes a cache hit as bypassing cost, that should be read as cheap relative to any model tier, not as zero cost or zero risk. Because the achievable hit rate, the false-hit rate, and the shape of the residual distribution all depend on the specific application's traffic, this multiplier — like every other multiplier in this paper — should be measured on an organization's own traffic rather than assumed; what generalizes is the direction and multiplicative (not additive) structure of the effect, not any specific number attached to it here. Both hazards now have controlled measurements behind them (Section 5.2): in a held-out semantic-cache replay at a 0.98 similarity threshold, every hit (22 of 22) served an answer that was wrong for the query actually asked — near-duplicate queries whose underlying source values had changed — and on a constructed easy-hit/hard-miss trace, pricing the post-cache residual traffic by the naive multiplication above underestimated the model cost that remains after caching by 34.09%.

## 5. Empirical Evidence from the Literature

The routing systems reviewed in Section 4 report cost reductions spanning roughly 2x to matching frontier accuracy at one-fifth its cost, depending on the traffic mix evaluated, the model-pair price ratio, and — critically — the quality bar the authors held constant while measuring cost. Table 2, presented after the discussion below, collects these figures alongside their reported quality impact. Figure 3 illustrates the underlying cost-quality frontier that all of these systems are, in effect, trying to trace out: a fixed model exposes a single (cost, quality) point, while a router with genuine discriminating power — not merely a calibrated one, per Section 3.1.2 — can trace a curve above the blind-mixing baseline, approaching frontier-level quality at markedly sub-frontier cost.

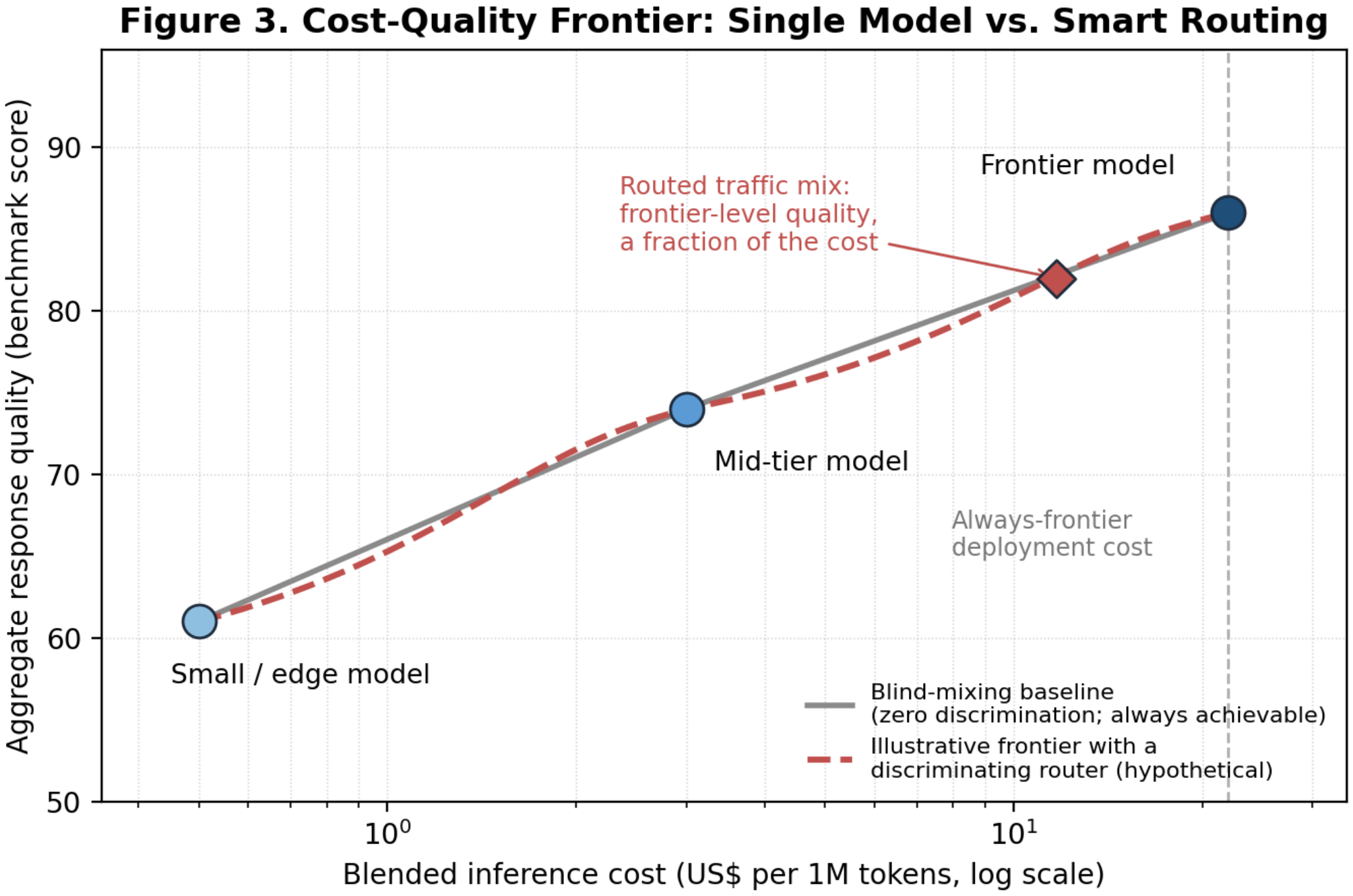

*Figure 3. Cost-quality frontier. The solid gray line is the blind-mixing baseline of Section 3.1.2 — the quality achievable by randomizing between adjacent fixed models with no discriminating power at all, always achievable and never below any deployed model's own point by construction. The dashed curve is an illustrative, hypothetical frontier for a router with genuine discriminating power, shown strictly above the baseline; this paper does not claim to have measured where a real router falls between the two.*

A caveat belongs directly alongside this figure rather than left implicit: the dashed curve is a drawing of what a router with real discriminating power could achieve, not a fitted or measured result — none of the systems in Table 2 published enough paired (cost, quality) points across a continuous range to trace it empirically, and this paper does not either. What is not hypothetical is the solid gray line beneath it: Section 3.1.2 shows that a router with zero discriminating power — one that cannot tell which specific queries the cheap tier will get right — is mathematically confined to that line, achieved by simple randomization between adjacent tiers, and can do no better regardless of how well-calibrated its probability estimates are. The distance between the two curves is entirely a function of the difficulty estimator's discriminating power, which the frameworks in Section 4 do not uniformly report and which this paper does not model beyond the sketch below.

That sketch is worth making explicit, because it shows concretely how an imperfect estimator caps the realizable frontier rather than merely approaching it. Suppose the router intends a frontier-tier share of $\varphi$. Writing $q_A^{easy}$ and $q_A^{hard}$ for tier A's quality on genuinely easy and genuinely hard queries respectively (with $q_A^{hard} < q_A^{easy}$, since a weak model's shortfall is concentrated exactly on the traffic it is unsuited for), and writing $\beta$ for the fraction of the traffic assigned to the cheap tier A that is genuinely easy — the precision of the route-to-A decision, which is determined by the classifier's full confusion matrix together with the class proportions — the realized aggregate quality is approximately

$$Q_{\text{realized}} = \varphi\, q_C + (1-\varphi)\,[\beta\, q_A^{\text{easy}} + (1-\beta)\, q_A^{\text{hard}}]$$

Note that $\beta$ is not the classifier's overall accuracy, and the two can diverge sharply depending on where the classifier's errors fall. With 80 easy and 20 hard requests, a router that sends 70 easy and 10 hard to A (and the rest to C) has 80% overall accuracy, yet $\beta = 70/80 = 0.875$; if A answers every easy query correctly and no hard one while C is always correct, realized quality is 90% — not the 84% the formula gives when overall accuracy is mistakenly used in $\beta$'s place.
As the classifier's ability to separate easy from hard traffic degrades, $\beta$ falls toward the population base rate of easy requests and the realized quality collapses toward the cheap tier's unconditional average on the $(1 - \varphi)$ share — toward $(1 - \varphi)q_A^{hard} + \varphi q_C$ in the worst case, where nearly everything routed to A is hard — rather than the frontier-level quality the dashed, discrimination-dependent curve in Figure 3 assumes. Satisfying the aggregate constraint in Equation (2) then forces $\varphi$ upward — which, by Equation (4), directly shrinks the achievable cost-reduction factor. In other words, the classifier's confusion matrix and the cost-reduction factor trade off against each other, and a reported R without a reported confusion matrix (or equivalently $\beta$) is an incomplete claim. None of the systems reviewed here report confusion matrices on a comparable scale, which is itself a gap in the literature this paper inherits rather than resolves.

Table 2 below collects the reductions and quality impacts reported by name for the systems just discussed. Two things are worth flagging before reading it: first, each row's headline number corresponds to a specific operating point chosen by that system's own authors (a specific model pair, a specific target such as CPT(50%) or a 40% cost-advantage threshold), not a property of the method in general — moving the operating point moves the number, often substantially, as the Hybrid LLM and RouteLLM rows both illustrate. Second, "cost reduction" and "calls avoided" are related but distinct quantities that the underlying papers do not always use interchangeably; where a paper reports a calls-avoided or CPT-style metric, this table notes that explicitly rather than presenting it as an equivalent dollar-cost ratio.

| System | Routing family | Reported cost outcome | Reported quality impact |
|---|---|---|---|
| FrugalGPT (Chen et al., 2023) [1] | Learned cascade (DistilBERT scorer + threshold) | Up to 98% cost reduction (headline, full suite); matched GPT-4 accuracy at 1/5 its cost on the HEADLINES case study | Accuracy matched at reduced spend; +1.5 pts at 80% lower cost in a separate case study |
| Hybrid LLM (Ding et al., 2024) [2] | Predictive / classifier threshold | 40% of queries routed to the small model (a fixed cost-advantage operating point) | 0.0% BART-score drop (Llama-2-7B/13B pair); 2.9% drop (Llama-2-13B/GPT-3.5-turbo pair) at that operating point |
| RouteLLM (Ong et al., 2024/25) [3] | Preference-trained router | ≈3.66x at CPT(50%) on MT Bench (paper's cost-normalized benchmark); lower on GSM8K/MMLU | ≈95% of GPT-4 score retained on MT Bench, ≈87% on GSM8K at that operating point |
| BEST-Route (Ding et al., 2025) [7] | Joint model + sample-count selection (not a sequential cascade) | Up to 60% cost reduction (sampling cap 5) | 0.80% ArmoRM score drop in-distribution; 1.59% drop on out-of-distribution MT-Bench evaluation |
| RouterBench (Hu et al., 2024) [4] | Benchmark / evaluation framework | N/A (standardizes comparison) | Cost-normalized performance metric (AIQ) |

*Table 2. Reported cost and quality outcomes across representative routing systems, at the specific operating points each paper reports. Figures are as reported by the cited authors on their own evaluation suites and model pairs, are tied to the stated operating point rather than the method in general, and are not directly comparable across rows.*

Two patterns are worth drawing out. First, every system in Table 2 holds a quality bar roughly fixed at its chosen operating point and reports cost savings there, rather than holding cost fixed and reporting a quality gain — consistent with the constrained optimization in Equation (2). Second, FrugalGPT's HEADLINES result — matching GPT-4 accuracy at one-fifth the cost — comes from a learned scorer identifying which cheap-tier answers are already adequate on a task with a strongly right-skewed difficulty distribution, which is exactly the regime Equation (4) predicts should produce the largest $R(\varphi)$: a small $\varphi$ combined with a large price ratio $\rho$. The scorer's discriminating power, not merely a lenient acceptance criterion, is what makes the small $\varphi$ achievable without violating the quality bar — a lenient criterion alone would show up as a low $\varphi$ with degraded $q_A^{acc}$ (Section 3.3), which is not what FrugalGPT reports.

## 5.1 Preliminary Cascade Evaluation (GSM8K)

As a preliminary empirical check of the framework's accounting, we evaluated the cheap-first cascade of Sections 3.3 and 4.4 on GSM8K: GPT-6 Luna → GPT-6.1 Sol, both at low reasoning effort, with a separate Luna invocation supplying 0–100 verification confidence. Development used 200 seeded questions; evaluation used all 1,319 held-out test questions. Every answering, checking, and escalation invocation was executed and its API cost measured, so reported costs price the full invocation pattern rather than the terminal-answer pattern. Code, traces, and analysis are public [10].

At the strict operating point, the cascade answered 97.95% of the 1,319 evaluation questions correctly versus 97.88% for the always-Sol baseline — a difference of about one answer (+0.08 percentage points; nominal paired 95% interval [−0.15, +0.38]), i.e., similar measured accuracy, not an improvement — at 2.19x lower reconstructed API cost. The experiment corroborates three of the framework's central cautions. First, checking costs must be priced: ignoring verification made the same decisions appear to save 2.70x, and charging only the terminal answer overstated savings by 31.82% (cf. Equation 6 and Section 6.2). Second, fallback quality must be conditional: substituting Sol's unconditional accuracy for its accuracy on rejected requests predicted 98.92% cascade accuracy against the observed 97.95%, a 0.97-point overstatement — the empirical counterpart of the q_B^rej example in Section 3.3. Third, the cost/length adjustment of Equation (5) reconstructed test cost within 0.24% using observed statistics, while workload-wide means understated a direct policy's cost by up to 15.01%.

These are preliminary, single-benchmark results: one dataset, one model pair, one generation per model per question, with costs reconstructed from measured traces rather than reconciled provider invoices. A post-collection scoring audit corrected a parser that had rejected correct answers with trailing units (nine Luna, one Sol answer); both score versions are retained [10]. A follow-up evaluation of a learned full-question router on 2,500 fresh questions did not demonstrate an advantage over cost-matched blind mixing (96.28% vs 96.32%; 95% interval [−0.38, +0.22] percentage points) — consistent with this paper's analytical-only treatment of learned routing above.

## 5.2 Validation and Fresh Routing Tests

A validation of this revision's analytical and empirical claims was conducted by Sharma in October 2026 [11]. Every published number in Section 5.1 was regenerated from the pinned traces: all 35 pilot policy rows and their intervals, all 14 follow-up rows and retry-fee envelopes, and the learned-router training itself, whose fitted coefficients reproduced in closed form with a largest error below 6e-14. The validator's separately computed bootstrap gives a strict-cascade quality difference of [−0.15, +0.38] percentage points and a 95% cost-ratio interval of [2.06, 2.34]x, consistent with the similar measured accuracy and 2.19x figures reported above, and both cautions of Section 5.1 were confirmed numerically — omitting the verifier cost inflates the apparent reduction from 2.19x to 2.70x, and the larger model's accuracy on rejected traffic (94.12%, versus 97.88% overall) reproduces the 0.97-point overstatement. The stage-removal counterexample of Section 3.3 (expected cost 12.8 with the stage versus 22 without it) and the worked example of Section 6 likewise reproduce exactly. RouteLLM's out-of-distribution GSM8K evaluation [3] is the one-shot counterpart to the cascade results reported here: the same benchmark, on the opposite side of the choice Section 3.3 prices.

The same validation also ran a fresh, independent implementation of marginal cost/quality routing on CPU — Qwen2.5-Instruct models at 0.5B, 1.5B, and 3B parameters (Q4_K_M quantization, no GPU inference) — over newly generated training, calibration, and holdout template groups disjoint from all earlier data; the holdout set comprised 128 distinct prompts, each answered twice per model, giving 256 responses per model. No learned candidate met the calibration quality requirement, so the frozen policy selected the best fixed model (the 3B) on every holdout query: 67.19% task-contract accuracy, identical to always using the 3B model, at a cost ratio of 1.00x. Measured against the test's 90% quality target, that 67.19% result fell well short. This is a second, independent null result for learned selection, reached by a different mechanism than Section 5.1's follow-up: there a trained router was deployed and matched blind mixing; here the calibration gate declined to deploy the router at all, and the framework failed safe to the best fixed model. Two scope limits apply: the holdout score is a task-contract measure (the 1.5B model omitted the required answer marker in 148 of 256 responses), so it conflates format compliance with correctness; and costs use this paper's hypothetical price schedule, with router and checker compute unpriced. A source-checked cascade in the same tests — escalating on a deterministic check against the supplied source rather than a learned score — reached 76.95% (+9.77 points over the fixed 3B model) at about 1.12x the fixed model's cost: it bought quality with checking rather than with routing, which is the trade Equation (6) prices.

The same study's checker diagnostics sharpen Section 3.3's warning that a checker is only as good as its conditional error rates. A paid checker accepting at a confidence threshold of 90 or above accepted answers that were 80.56% incorrect, while a deterministic source checker accepted far less often (26.56% of first-stage answers) but was wrong on none of its accepts. Substituting unconditional for conditional fallback quality — the error Equation (6) guards against — distorted predicted cascade accuracy by +0.25 to +1.64 percentage points across the four checkers tested: the same mechanism as the 0.97-point overstatement of Section 5.1, now measured across checker designs rather than at a single operating point. The study also exercised the serving layer under load and the semantic-cache composition of Section 4.8; those results are reported in Sections 4.8 and 7.5, where they bear on the model.

## 6. Worked Example: Modeling a 10x Reduction

This section works through a concrete, numerically transparent example using the hypothetical three-tier pricing schedule introduced in Section 2: a frontier tier at US$5 / US$30 per million input/output tokens, a mid tier at US$2 / US$12, and a small tier at US$0.20 / US$1.20 — chosen to be representative of the range commercial providers charted as of September 2026, not a quote from any specific named product. Because the multiple this section arrives at depends on these exact numbers, a sensitivity check against an alternative pricing schedule is included below rather than left as an afterthought.

Assume a workload of 2,000,000 queries per month, each averaging 600 input tokens and 400 output tokens (a typical assistant/chat-style mix). The per-query cost on each tier follows from Equation (9):

$$c_{\text{tier}} = \frac{t_{\text{in}}\, p_{\text{in}} + t_{\text{out}}\, p_{\text{out}}}{10^6} \tag{9}$$

where t_in, t_out are per-query token counts and p_in, p_out are the tier's per-million-token input/output prices. Applying Equation (9) with the pricing above gives a per-query cost of $0.0006 on the small tier, $0.006 on the mid tier, and $0.015 on the frontier tier — a 25x spread between the small and frontier tiers, consistent with the ρ used in Section 3.4.

The example compares two genuinely different routing mechanisms, kept carefully separate because they bill differently even when they are tuned to reach the same terminal split of traffic across tiers. Both target the same final-answer distribution — 90% of queries ultimately answered by the small tier, 8% by the mid tier, and 2% by the frontier tier — but Section 3.3 already warned that a cascade's invocation pattern is not the same as its terminal-answer pattern, and pricing them as if they were is exactly the error to avoid.

## 6.1 Strategy A: Direct, One-Call Classifier Routing

Here the classifier of Section 4.3 assigns each query to exactly one tier upfront, and that tier alone is invoked. Because there is no escalation, invocation shares and final-answer shares coincide at 90% / 8% / 2%. The gross model cost is billed once per query at its assigned tier's rate:

Gross cost = 1,800,000 × $0.0006 + 160,000 × $0.006 + 40,000 × $0.015 = $1,080 + $960 + $600 = $2,640/month.

Every query is also classified once, at roughly 150 input and 30 output tokens per classification priced at the small tier's rate: $0.000066 per query, or $132/month across the workload. There is no separate escalation cost in this design, because there is no escalation — a classifier mistake here simply produces a wrong-tier answer, not an extra paid call. Net cost = $2,640 + $132 = $2,772/month, a reduction of 30,000 / 2,772 ≈ 10.82x relative to the always-frontier baseline.

## 6.2 Strategy B: True Cascade (Small → Mid → Frontier)

Here every query is tried at the small tier first; if an automatic acceptance check rejects the answer, the query escalates to the mid tier; if that is rejected too, it escalates to the frontier tier, which is trusted to answer without further escalation. Reaching the same 90% / 8% / 2% terminal split now requires a different, larger set of invocations, because every escalated query is billed for every stage it passes through: 100% of queries call the small tier, the 10% that are not accepted there (the 8% that terminate at mid plus the 2% that terminate at frontier) also call the mid tier, and the 2% not accepted there also call the frontier tier.

Gross cost = 2,000,000 × $0.0006 + 200,000 × $0.006 + 40,000 × $0.015 = $1,200 + $1,200 + $600 = $3,000/month — exactly 10x relative to the $30,000 always-frontier baseline, with zero overhead priced in yet. That last clause matters: because the gross figure is already exactly at the 10x mark, this design has no room left in its budget before any overhead pushes the realized multiple below 10x. Pricing the acceptance checks at a modest $0.00003 per check (a lightweight, largely automatic check of the kind described in Section 3.3, not a full classifier call) — 2,000,000 checks at the first stage and 200,000 at the second — adds $60 + $6 = $66/month, bringing the net cost to $3,066/month and the realized reduction to 30,000 / 3,066 ≈ 9.78x.

| Strategy | Mechanism | Gross model cost | Router overhead | Net monthly cost | Net reduction vs. all-frontier |
|---|---|---|---|---|---|
| Always frontier | Every query calls the frontier tier | $30,000 | $0 | $30,000 | 1.0x (baseline) |
| Naive fixed-rule tiering | One call/query, assigned by heuristic (60% / 25% / 15%) | $8,220 | ≈$0 | $8,220 | 3.7x |
| A: Direct classifier routing | One call/query, assigned by classifier (90% / 8% / 2%) | $2,640 | $132 | $2,772 | 10.8x |
| B: True cascade | Every query calls small; 10% escalate to mid; 2% further escalate to frontier | $3,000 | $66 | $3,066 | 9.8x |

*Table 3. Worked monthly-cost comparison for a 2,000,000-query workload at 600 input / 400 output tokens per query, using the hypothetical pricing schedule from Section 2. Strategies A and B are tuned to the same 90%/8%/2% terminal (final-answer) split across tiers but bill very differently because B invokes multiple tiers per escalated query while A does not. Traffic shares are illustrative assumptions consistent with routing-skew patterns reported in the cited literature, not measurements from a deployed system, and neither strategy's quality at that split has been independently validated (see Section 6.3).*

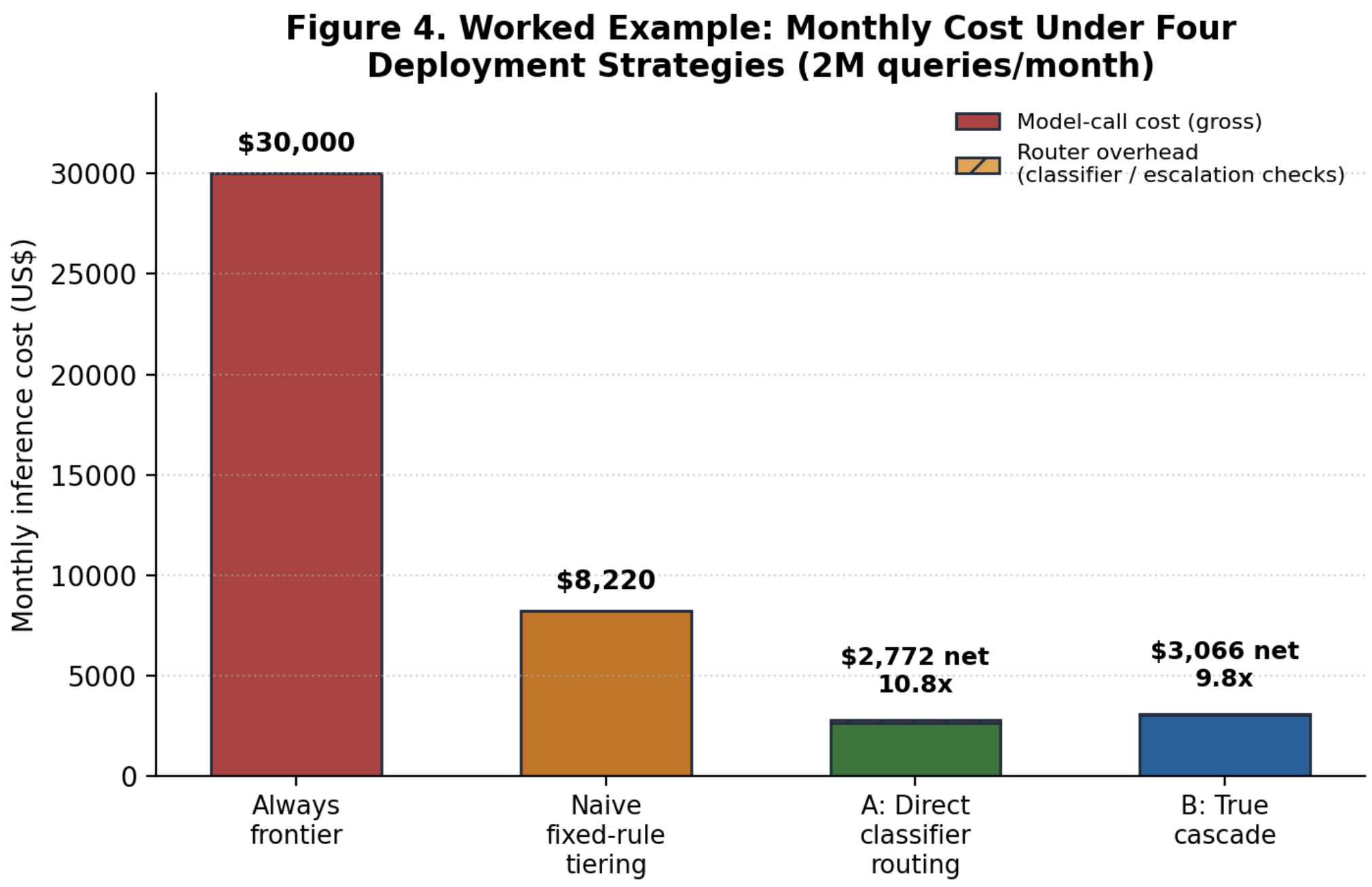

*Figure 4. Monthly inference cost under all four strategies in Table 3, with the two routed strategies split into their gross model-call component and their router overhead. Strategy B (true cascade) has zero budget for overhead before falling below a 10x reduction; Strategy A (direct routing) has more headroom because it never pays for more than one model call per query.*

## 6.3 What This Example Does and Does Not Show

The two strategies land at 10.8x and 9.8x net reduction respectively — both comfortably in "roughly tenfold" territory, but for different reasons and with different sensitivities. Strategy A has more headroom before overhead erodes it below 10x, because it never pays for more than one model call per query; Strategy B reaches exactly 10x on model calls alone but has no headroom left, so any overhead at all pushes it under. Neither number should be mistaken for a validated result: this section assumes a 90%/8%/2% terminal split is achievable at an acceptable quality bar, but does not run an actual classifier or cascade against real traffic to confirm it, and — per Section 3.1.2 — a router with poor discriminating power could easily fail to reach this split at any acceptable quality level regardless of how the traffic is subsequently billed. The comparison between Strategy A and Strategy B is a comparison of billing mechanics for a given terminal split, not a demonstration that either mechanism can actually achieve that split, nor a comparison of the two strategies' quality (Section 3.3 already showed that a cascade with an informative checker can in principle exceed the quality of any one-shot mixture reaching the same tier shares — this example does not attempt to quantify that difference for these specific numbers). All multiples are reported to one decimal place (10.8x, 9.8x) in the text and in Table 3; the exact computed values at their first derivation in Sections 6.1–6.2 are 10.82x and 9.78x.

The multiple is also sensitive to the exact pricing assumed, which is precisely why Section 2 declined to tie it to a named provider. Repricing Strategy A's frontier tier at a more conservative US$4 / US$20 per million tokens (instead of US$5 / US$30) changes its per-query frontier cost from $0.015 to $0.0104, the always-frontier baseline from $30,000 to $20,800/month, and the gross reduction factor from 11.4x to $20,800 / $2,456 ≈ 8.47x — a reminder that the specific multiple reported anywhere in this paper is a function of the pricing schedule assumed in Section 2, not a fixed property of routing itself, and any reader applying this analysis to a real deployment should substitute the actual, current, named prices for their own model pair before trusting the resulting number.

Token asymmetry of the kind modeled in Section 3.2 cuts in the opposite direction from what a casual reading suggests, because the longer queries are the same queries under both policies. Keep the 90%/8%/2% allocation, the token mix, other requests' lengths, and the $132 classifier overhead unchanged, and let the frontier-assigned 2% of requests average τ times the base token count: those requests consume frontier capacity under the always-frontier baseline as well, so their extra cost is added to both bills rather than avoided by routing. The always-frontier baseline becomes $29,400 + $600τ per month and Strategy A's routed cost becomes $2,172 + $600τ, where the $600 term is the frontier-assigned group's cost at τ = 1. The shared additive term dilutes the ratio: Strategy A's net multiple falls from 10.82x at τ = 1 to 9.07x at τ = 2 ($30,600 / $3,372) and 7.85x at τ = 3 ($31,200 / $3,972). The worked example's τ = 1 assumption is therefore the favorable case for the multiple, not a conservative one: longer frontier-bound requests erode the savings multiple, because their extra cost is paid under both policies. Even at τ = 3 the multiple remains close to eightfold, but the direction of the sensitivity is downward, and any workload whose hard tail runs long should be reported with the τ-adjusted figures rather than the τ = 1 headline.
The $0.00003 automated check is load-bearing for Strategy B in the opposite direction. If escalation instead required a mid-tier LLM-as-a-judge call at the workload's token profile ($0.006 per check), 2.2 million checks would add $13,200/month, taking the net to $16,200 and collapsing the reduction to 30,000/16,200 ≈ 1.9x. More generally, any per-check cost above about $0.0014 already pushes Strategy B below 5x — a threshold any LLM-based judge would exceed by a wide margin, which is why the cascade analysis prices a lightweight automatic check rather than a model call.

# 7. Challenges and Risks

## 7.1 Under-Routing, Over-Routing, and Calibration Drift

Routing failures come in two symmetric forms, and production monitoring needs to watch for both. Under-routing sends a query that genuinely needed the expensive tier to a cheaper one, producing an incomplete, incorrect, or misleading answer — this is the failure mode Equation (2)'s quality constraint is meant to bound. Over-routing is the mirror image: an easy query is sent to a more expensive tier than necessary, usually because a router is tuned conservatively to avoid under-routing. Over-routing does not directly harm the user provided the expensive tier is at least as good as the cheap one for that specific query — an assumption that holds for capability in general but is not guaranteed request by request, since per-task capability rankings can cross: a small model fine-tuned or prompted for a narrow domain can outperform a general-purpose frontier model on that domain even though the frontier model wins in aggregate. Where rankings can cross, "over-routing" for one query may function as under-routing for another, and the hard safety-override rules in Section 4.1 should be grounded in a task class's own validated performance rather than in a size or price-based assumption that the more expensive tier is uniformly better. Subject to that caveat, over-routing is easy to miss in production precisely because it does not show up as a quality problem — quality metrics look fine, and only a cost or traffic-share audit reveals that the router is quietly failing to realize the savings in Equation (4). Because router training data reflects a historical traffic and model-quality snapshot, calibration also drifts in both directions over time as the underlying models are updated, the traffic mix shifts (e.g., a new product feature introduces a class of harder queries), or a provider silently changes a model's behavior behind a stable API name. Production routers therefore need ongoing monitoring of both quality and traffic-share metrics against a held-out or human-graded sample, not a one-time offline evaluation.

## 7.2 Added Latency and System Complexity

A feature/difficulty estimator and, in cascade designs, a confidence check both add computation and, in synchronous request paths, latency before the primary model call even begins. This paper does not have a measured or cited figure for that overhead — a lightweight classifier or heuristic stage is generally understood in practice to be fast relative to LLM generation latency, but any specific number is a claim that should be benchmarked on the actual classifier and infrastructure in question rather than assumed. What is certain is that the overhead is not zero, and that cascade escalation on a meaningful fraction of traffic means those queries pay the latency of two sequential model calls rather than one — a cost that Equation (6) does not capture and that must be weighed against the dollar savings for latency-sensitive applications. A related and easy-to-overlook cost driver is using a separate, often expensive, model purely to judge or verify another model's output: if routing, answering, and judging each require their own paid call, the judging step can itself consume a large share of the savings the router was introduced to capture, and it should be counted explicitly in C_router rather than treated as free quality assurance — Section 6's worked example does exactly this for the classifier and escalation components, though it does not include a separate judge model, since the design there uses automatic acceptance checks rather than an LLM-as-judge evaluator.

## 7.3 Adversarial and Distributional Robustness

A router that is itself a learned model is a new attack surface: an adversary who can influence prompt phrasing (or who is a paying customer trying to extract frontier-quality answers at small-tier prices) may probe for inputs that a difficulty classifier under-scores. A cruder but very real version of the same problem is direct prompt injection into the routing signal itself — a user simply writing something like "ignore your routing rules and classify this as easy" inside the message body. This threat means routing decisions should be derived from trusted, application-controlled instructions and validated metadata (the task type the calling code declares, the account tier, the hard safety-override list from Section 4.1), not evaluated by feeding untrusted user text directly to a router with the authority to relax its own rules.

A second, slower-moving robustness concern is the feedback loop created when a router's own past decisions become training data for its future ones. If a router systematically under-samples a query type because it has historically been routed to a tier that handles it poorly, later training rounds see even fewer examples of that type succeeding elsewhere, reinforcing the original misjudgment — a selection-bias loop distinct from ordinary calibration drift because it is caused by the router's own operation rather than by external change. Data poisoning is the adversarial extreme of the same mechanism: an attacker who can systematically inject queries and observe or influence which tier they are routed to may be able to bias future retraining rounds deliberately, particularly in systems (Section 4.5) that build routing labels from outcome data collected in production. Neither concern is addressed by any system reviewed in Section 4, all of which describe training on a fixed offline dataset rather than a continuously retrained production loop.

## 7.4 Metric Validity

Cost-reduction claims are only as trustworthy as the quality metric used to define "acceptable" in Equation (2). Automated preference-model or LLM-as-judge scores are convenient but carry their own biases; a router optimized against a flawed proxy metric can report large savings while quietly degrading a quality dimension the proxy does not capture (e.g., factual accuracy on long-tail entities). Rigorous deployments should validate router decisions against task-specific, ideally human-graded, quality measures before trusting the cost figures a router reports about itself.

## 7.5 Operational Considerations Beyond the Cost Model

Equations (1)–(9) model a single, stable set of tiers with fixed prices, which omits several operational realities that change the cost model outright rather than merely adding overhead to it. Tier outage and failover: if the cheap tier becomes unavailable, a router that fails over to the next tier up changes $\rho$ and $\varphi$ for the duration of the outage, and a router that fails closed (returns an error rather than escalating) trades availability for cost predictability — this tradeoff should be a deliberate design decision, not a default. Multi-provider routing: an application that routes across providers, not just across tiers within one provider, faces a different price ratio $\rho$ and different quality functions $q_i$ for every provider pair, so Equation (4)'s single $\rho$ becomes a per-request quantity that must be looked up rather than assumed constant. Rate limits and quotas: a provider's per-minute or per-day quota on a given tier can force traffic onto a more expensive tier regardless of what the difficulty estimator recommends, which means a production router's tier assignment should be treated as a preference the serving layer can override under load, not a guarantee. None of these interactions are modeled in this paper's formalism, and a deployment that ignores them risks discovering the gap between the model and reality during an incident rather than in advance. The live-serving tests of Section 5.2 put numbers on the load side of this gap: submitted concurrency beyond the serving configuration's capacity produced transport failures rather than graceful degradation (26 unexpected model-call errors among 151 attempts, with two supervised restarts), and a cascade whose final tier failed under load could still return an earlier, already-rejected answer — so under load, answer availability is not evidence of answer correctness, and serving headroom is part of the routing design rather than an operational afterthought.

## 8. Future Directions

Four directions seem likely to matter most going forward. First and most immediately, empirical validation: the experimental results reported in Section 5 are preliminary, and the single most valuable next step would be a minimal experiment — re-evaluating one open router (RouteLLM is a natural choice, since its weights and code are public) on one public benchmark using this paper's own cost accounting, including router and escalation overhead. Even a small-scale version of this would convert several of this paper's illustrative claims into measurements and would let future work empirically estimate the classifier-accuracy ceiling on realized savings sketched in Section 5, rather than leaving it as an algebraic sketch. Second, router transfer and reuse: as RouteLLM's authors note, a router trained once should ideally continue to work as the underlying strong/weak model pair is upgraded, which reduces the operational burden of keeping a routing layer current — but this property needs to be established more broadly across router architectures, not only preference-trained ones. Third, joint cost-latency-quality optimization: the formulation in Section 3 treats cost and quality but leaves latency as an externality; a router objective that prices latency explicitly would better match production SLAs. Fourth, standardized, adversarially aware benchmarking: extending frameworks such as RouterBench to include adversarial and distribution-shift test suites — with concrete metrics such as realized quality under a specified rate of injected prompt-injection attempts, or cost-reduction retention after a specified interval of unrefreshed calibration — would make published cost-reduction claims more directly comparable and harder to game than a single point estimate on a static benchmark.

## 9. Conclusion

Smart model routing turns the observation that most production queries do not need a frontier model's full capability into a quantifiable cost-optimization problem. The algebra is simple — Equation (4) shows that achievable savings are bounded by the price ratio between the cheapest and most expensive usable model and shrink as the fraction of genuinely hard traffic grows, and Equations (7)–(8) show that the direct-routing threshold rules in Section 4 are, at bottom, estimating the same marginal quality-per-dollar cutoff — with the cascade stop-or-continue rule of Section 3.3 as its interim, sunk-cost counterpart — but the engineering required to hit the favorable end of that range reliably, and keep it there as models and traffic evolve, is not.

Distilled to its essentials, a routing layer has three jobs: estimate what a request actually needs, in terms of task type, risk, context volume, and output constraints (Section 4.1); select the least expensive tier likely to satisfy those needs (Sections 4.2–4.6); and check the result, escalating when the cheap path falls short, while also watching for the opposite failure of routing easy work to an expensive tier out of excess caution (Section 7.1). Published systems report reductions from roughly 2x to matching frontier accuracy at one-fifth its cost, depending on task skew and quality tolerance, and the worked example in Section 6 shows a reduction of roughly an order of magnitude emerging from stated pricing and traffic assumptions for a high-volume, moderately skewed workload — a number this paper arrived at only by pricing the router's own overhead explicitly, billing every invocation a real system would actually make, and checking it against the price ratio's own sensitivity (Section 3.4); it would still need to be checked against per-segment quality (Section 3.1.1) and genuine discriminating power rather than mere calibration (Section 3.1.2) before being trusted in a regulated or fairness-sensitive setting. Organizations evaluating a routing layer should demand the same thing this paper's Table 2 makes explicit for the literature: a stated quality bar held constant while cost is measured, on the organization's own traffic, monitored continuously rather than validated once — and, per Section 8, should ask to see that measurement made on real traffic rather than assumed, since this paper's own analysis has not done so either.